\documentclass[12pt]{iopart}
\pdfoutput=1
\eqnobysec
\usepackage{graphicx}
\usepackage{amssymb}
\usepackage{color}
\def\keywords#1{\vspace{10pt}
     \begin{indented}
     \item[]\rm Keywords: #1\par
     \end{indented}}

\def\exp{\mathrm {exp}}

\def\be{\begin{equation}}
\def\ee{\end{equation}}
\def\bea{\begin{eqnarray}}
\def\eea{\end{eqnarray}}

\def\PRA{{\it Phys. Rev. A }}
\def\PRB{{\it Phys. Rev. B }}
\def\PRE{{\it Phys. Rev. E }}
\def\JSM{{\it J. Stat. Mech. }}
\def\EPL{{\it Europhys. Lett. }}

\def\bibspace{\hspace*{-3.5mm}}

\begin{document}
\jl{1}

\title[Evolution of the magnetization after a local quench]{Evolution of the magnetization after a local quench in the critical transverse-field Ising chain}

\author{Ferenc Igl\'oi$^{1,2,3}$, Gerg\H o Ro\'osz$^{1,2}$ and Lo\"{\i}c Turban$^{3,4}$}

\address{$^1$ Wigner Research Centre, Institute for Solid State Physics and Optics,
H-1525 Budapest, P.O.Box 49, Hungary}

\address{$^2$ Institute of Theoretical Physics,
Szeged University, H-6720 Szeged, Hungary}

\address{$^3$ Universit\'e de Lorraine, Institut Jean Lamour, UMR 7198, Vand\oe uvre l\`es Nancy Cedex, F-54506, France} 

\address{$^4$ CNRS, Institut Jean Lamour, UMR 7198, Nancy Cedex, F-54011, France}

\ead{igloi.ferenc@wigner.mta.hu, gergo$\_$roosz@titan.physx.u-szeged.hu {\rm and} loic.turban@univ-lorraine.fr}

\begin{abstract}
We study the time evolution of the local magnetization in the critical Ising chain in a transverse field after a
sudden change of the parameters at a defect. The relaxation of the defect magnetization is algebraic and the corresponding exponent, which is a continuous function of the defect parameters, is calculated exactly.
In finite chains the relaxation is oscillating in time and its form is conjectured on the basis of precise numerical
calculations.
\end{abstract}


\submitto{J. Stat. Mech.}

%\date{\today}

%\tableofcontents
%\newpage

\section{Introduction}
\label{sec:introduction}
Non-equilibrium relaxation of different observables in a closed quantum system at $T=0$ after a sudden change of
parameters in the Hamiltonian is of recent interest, both experimentally (see~\cite{bloch} for a review) and theoretically
(see~\cite{Polkovnikov_11} for a review). From a theoretical point of
view, up to now, much attention has been paid to the problem of the \textit{global quench}, when parameters of the Hamiltonian
are modified uniformly in space~[3]--[44]. 
%\cite{barouch_mccoy,igloi_rieger,sengupta,Rigol_07,Calabrese_06,Calabrese_07,Cazalilla_06,Manmana_07,
%Cramer_08,Barthel_08,Kollar_08,Sotiriadis_09,Roux_09,Sotiriadis_11,Kollath_07,Banuls_11,Gogolin_11,Rigol_11,Caneva_11,Cazalilla_11,
%Rigol_12,Santos_11,Grisins_11,Canovi_11,Calabrese_05,Fagotti_08,Silva_08,Rossini_09,Campos_Venuti_10,Igloi_11,Rieger_11,Foini_11,
%Calabrese_11,Schuricht_12,Calabrese_12,blass,Essler_12,evangelisti_13,fagotti_13,pozsgai_13a,fagotti_essler_13,collura_13}.
In this case one is concerned with the functional form of the relaxation process, as well as the properties
of the stationary state, both for integrable and non-integrable quantum systems.

Another interesting process is the \textit{local quench}, when parameters are modified on a given site. Experimentally, this happens
during x-ray absorption in metals~\cite{x-ray}. Most of the theoretical studies in this field are concentrated on one-dimensional systems, in particular at quantum critical points. 

In this case, analytical results have been derived using conformal field theory
(CFT)~\cite{cc-07loc,stephan_dubail}.
In a continuum description the system evolves in a space--time region with coordinates $(x,t)$. 
The value of some local parameter, such as the strength of a coupling at $x=0$, changes from $\kappa_1$ before
the quench ($t<0$) to $\kappa_2$
after the quench ($t>0$). The expectation values of different observables are obtained using path-integral techniques.
CFT generally works for appropriate boundary conditions: $\kappa=0$ (uncoupled half chains)
or $\kappa=\infty$ (fixed local spin) and $\kappa=1$, i.e., uniformly coupled chain. 
For the dynamical entanglement entropy~[48]--[52], %\cite{ep-07,ekpp-08,iszl-09,sodano10,zamora14}, 
after changing from $\kappa_1=0$ to $\kappa_2=1$, a logarithmic increase is found~\cite{cc-07loc} which takes the universal form,
${\cal S}(t) = (c/3) \ln t + {\rm const}$, involving the central charge $c$ of the CFT. In a finite system of total length $L$ the
time dependence is periodic, having a sinusoidal form in terms of $t/L$~\cite{stephan_dubail}. Using CFT, predictions have been made about the behavior of the magnetization and different correlation functions: power laws in time and distance from the local quench are found, with exponents given by combinations of bulk and surface static scaling dimensions~\cite{cc-07loc}.
Numerical results obtained on specific models are in good agreement with these CFT predictions~\cite{Divakaran_11}.

Local quenches in non-conformally-invariant systems have been studied, too. When the quench is performed in the ordered phase, a semi-classical approach can be used~\cite{Rieger_11} which has been numerically tested on the transverse Ising chain (TIC)~\cite{Divakaran_11}.
For disordered systems, such as the random TIC, a variant of the strong disorder renormalization group method~\cite{im} has been used to predict the time evolution of the entanglement entropy~\cite{Igloi_12,Vosk_12}, as well as the related
full counting statistics~\cite{Levine_12}. At the random critical point both quantities have an ultra-slow time dependence:
${\cal S}(t) \sim \ln \ln t$, which has been confirmed by numerical calculations.

In this paper we study the time evolution of the local magnetization at the critical point of the TIC, 
after a general local quench, when the values of both the local coupling and the local transverse
fields are changed at $t=0$. It is known from exact calculations
that the static critical behavior near a local defect in the TIC is non-universal, i.e., the scaling dimension
of the local magnetization varies continuously with the defect strength, $x_i=x_i(\kappa_i)$~[58]--[63]. %\cite{bariev79,mccoy80,kadanoff81,brown82,peschel84,igloi93}. 
The problem has later been treated using $S$-matrix theory~\cite{delfino94}, conformal methods~[65]--[70]
%\cite{turban85,guimaraes86,henkel87a,henkel87b,henkel88,henkel89} 
and conformal field theory~[71]--[73].
%\cite{oshikawa96,oshikawa97,leclair99}.
Similarly, the ground-state~\cite{iszl-09}, [74]--[76]
%\cite{peschel_03,ep-10,pe-12}
and the dynamical~\cite{iszl-09,ep-12} entanglement entropy across
a defect involves a prefactor, the so-called effective central charge, which is also a function of the defect
parameters. It is therefore expected that the 
non-equilibrium critical relaxation of the defect magnetization also involves exponents which are non-universal.
Our goal in this work is to calculate their parameter dependence, as well as to study the functional form of the defect relaxation
in large, but finite systems.

The structure of the paper is the following. The model and the scaling predictions about the (imaginary and real) time dependence of the local magnetization is presented in section~\ref{sec:model-scaling}. Results of numerical investigations for initially ordered and initially
non-ordered defects are presented in section~\ref{sec:numerical} and discussed in the final section. The calculation of the defects static exponents is presented in the appendix.

\section{Model and scaling behavior}
\label{sec:model-scaling}

\subsection{Model}
\label{sec:model}

We consider a critical TIC of length $L$ with free boundary conditions and a defect at $L/2$. The Hamiltonian of the system may be generally written as  
\be\fl
{\cal H}_i\!=\!-\frac{1}{2}\!\left[\sum_{n=1}^{L-1}\!\sigma_n^x \sigma_{n+1}^x\!+\!(J_i\!-\!1)\sigma_{L/2}^x \sigma_{L/2+\!1}^x
\!+\!\sum_{n=1}^L\sigma_n^z\!+\!(h_{i1}\!-\!1)\sigma_{L/2}^z+(h_{i2}\!-\!1)\sigma_{L/2+\!1}^z\right],
\label{hamiltonian}
\ee
where the $\sigma_n^{x,z}$'s are Pauli spin operators. The index $i=1,2$ refers to the values of the transverse fields $h_{i1}$, $h_{i2}$ and the coupling $J_i$, before and after the quench at $t=0$. We are interested in the time dependence of the local magnetization after the quench, $m_n(t)$, which is given by the off-diagonal matrix element~\cite{Yang_52} of the magnetization operator in the Heisenberg picture, $\langle \Phi_0 | \sigma_n^x(t) | \Phi_1 \rangle$,
between the ground-state $|\Phi_0 \rangle$ of the initial Hamiltonian ${\cal H}_1$ and 
its first excited state $|\Phi_1 \rangle$.  In the Heisenberg representation, the magnetization operator is given, for $t>0$, by $\sigma_n^x(t) = e^{-i {\cal H}_2t} \sigma_n^x e^{i {\cal H}_2t}$ where  ${\cal H}_2$ is the Hamiltonian after the quench.
We follow the evolution of the local magnetization at the defect, $m_d(t)=m_{n=L/2}(t)$.

\subsection{Scaling behavior in imaginary time}
\label{sec:imaginary}

Let us first analyze the scaling behavior of the defect magnetization in
imaginary time $t=i \tau$ at criticality. Then the process viewed in the $n-\tau$ plane corresponds to a two-dimensional ($2d$) critical classical Ising model with a composite ladder defect at the center (see figure \ref{figA1} for an illustration; the $x$ ($y$) axes there correspond to $\tau$ ($n$) here). The parameters of the defect are different for $\tau<0$
and for $\tau>0$, respectively. Along the defect line the magnetization has the finite-size scaling behavior
\be
m_d(\tau,L) = L^{-x_i}\, \widetilde{m}^i_d(\tau/L),\quad i=1(2),\quad \tau<0\  (>0)\,.
\label{m_d}
\ee
According to exact calculations, which are recapitulated in the appendix, the
local scaling exponent $x_i$ depends on a combination of the defect parameters
\be
\kappa_i=\frac{J_i}{h_{i1}h_{i2}}\;.
\label{kappa_ii}
\ee
and it is given by
\begin{equation}
x_i=\frac{2}{\pi^2} \arctan^2\left(\frac{1}{\kappa_i}\right)\,.
\label{x_d}
\end{equation}
The scaling function $ \widetilde{m}^i_d(z)={\rm const}$ for $|z| \gg 1$, i.e., for $|\tau| \gg L$ and for $|z| \ll 1$
it has a power-law dependence $ \widetilde{m}^i_d(z) \sim |z|^{\omega_i}$. The value of the exponent $\omega_i$ is related
to the scaling behavior of the local magnetization in the region $\tau\ll L$, where the two different semi-infinite defect lines meet.
Here the local critical behavior is influenced by both defects and, asymptotically, we have
\be
m_d(\tau\ll L,L) \sim L^{-x_{12}}\,,
\label{m_d_0}
\ee
where $x_{12}$ is the composite defect (or generalized corner) exponent. As shown in the appendix,
$x_{12}$ is given by the geometric mean of $x_1$ and $x_2$ (see equation~\eref{eA12}). 

The scaling behaviors in equations~\eref{m_d} and~\eref{m_d_0} remain compatible if the exponent $\omega_i$ takes the form $x_{12}-x_i$. As a consequence the magnetization profile for $0<\tau \ll L$ behaves as
\begin{equation}
m_d(\tau) \sim \tau^{x_{12}-x_2}\,,\quad 0<\tau \ll L\,.
\label{m_tau}
\end{equation}

Let us have a few comments before closing this section.
First, the scaling behavior is the same on both sides of the defect.
Second, the fixed-spin initial condition can be realized either with $h_{1j}=0$ ($j=1$ and/or $2$) or with $J_1=\infty$, leading to $\kappa_1=\infty$. Then, according to~\eref{x_d}, $x_1=x_{12}=0$, so that~\eref{m_tau} simplifies to
\begin{equation}
m_d^{(+)}(\tau) \sim \tau^{-x_2}\quad 0<\tau \ll L\;.
\label{m_+}
\end{equation}
Our third and final comment is about a protocol where two half-chains,
initially disconnected, are connected by a bulk coupling for $\tau>0$. Let $x_m$ and $x_{ms}$ be the scaling dimensions of the magnetization, in the bulk and at a free surface, respectively. Thus initially $\kappa_1=0$ and $x_1=x_{ms}=1/2$  whereas $\kappa_2=1$ and $x_2=x_m=1/8$ for $\tau>0$. In this case the magnetization behaves as
\be
m_d^{(fb)}(\tau)\sim \tau^{1/8}\,,\quad 0<\tau \ll L\,.
\label{m_fb}
\ee
\subsection{Scaling behavior in real time}
\label{sec:real}
Concerning the scaling behavior of the magnetization in real time, some results have been obtained in special cases, when in the initial state either the spin at the defect
is fixed ($\kappa_1=\infty$) or the chains are disconnected (free) ($\kappa_1=0$), and the final state is
the homogeneous bulk one ($\kappa_2=1$).

For fixed-spin initial state the local magnetization has been predicted by CFT to decay as~\cite{cc-07loc}
\begin{equation}
m_d^{(+)}(t) \sim t^{-2 x_m}\quad 0<t \ll L\,,
\label{m_+t}
\end{equation}
a result which has been checked numerically on the TIC~\cite{Divakaran_11}. In a finite system, for large $t$ and $L$, the time dependence of the local magnetization is periodic. In an open chain the numerical
results are well described by the following sinusoidal form~\cite{Divakaran_11}
\begin{equation}
m_d^{(+)}(t,L) \sim \left[L\sin\left(\pi \frac{t}{L} \right)\right]^{-2x_m}, \quad 0<t < L\,,
\label{m_+tL}
\end{equation}
which reduces to~\eref{m_+t} for $t \ll L$.

For the initial state with two disconnected chains, the numerical results on the TIC are summarized in the following
conjectured formula~\cite{Divakaran_11}:
\begin{equation}
m_d^{(fb)}(t,L) \sim L^{-1/2}\left[L\sin\left(\pi \frac{t}{L} \right)\right]^{1/4}, \quad 0<t < L\,.
\label{m_fbtL}
\end{equation}
At short time it behaves as
\begin{equation}
m_d^{(fb)}(t) \sim m_0(L)\,t^{1/4}, \quad 0\leq t \ll L\,,
\label{m_fbt}
\end{equation}
where $m_0(L) \sim L^{-x_{ms}}$ is the equilibrium value of the defect magnetization in the initial state.

In equation~\eref{m_fbt} the time exponent can be rewritten in the form $1/4=2(x_{ms}/2-x_m)$ where
the exponents $x_{12}=x_{ms}/2$ and  $x_2=x_m$ of the protocol ($\kappa_1=0$ to $\kappa_2=1$) are in evidence.
The same form applies to~\eref{m_+t} where $x_{12}=0$ for a fixed-spin initial state.
Comparing the scaling behaviors in imaginary time (equations~\eref{m_+} and \eref{m_fb}) and real time (equations~\eref{m_+t} and \eref{m_fbt}), we see that they are equivalent if one substitutes $t^2$ for $\tau$.
Thus we conjecture that in real time the magnetization generally
behaves as
\be
m_d(t) \sim m_0(L)\,t^{2(x_{12}-x_2)}, \quad 0<t \ll L\,,
\label{m_t}
\end{equation}
with $m_0(L) \sim L^{-x_1}$.
The periodic behavior in a finite-size system is then expected to be given by
\begin{equation}
m_d(t,L)\!\sim\! L^{-x_1}\left[L\sin\!\left(\pi \frac{t}{L} \right)\right]^{2(x_{12}-x_2)}\!\!\!\!\!\!,~ 0<t < L \,.
\label{m_tL}
\end{equation}
In the following, the validity of the two last equations, which constitute our main  result, 
will be checked through large scale numerical calculations.

\section{Numerical investigations}
\label{sec:numerical}
\subsection{Technical details}
\label{sec:technic}

In the numerical investigations we make use of the fact that the TIC can be expressed in terms of free
fermions~\cite{pfeuty70}. The local magnetization is given by a Pfaffian~\cite{schultz64} which is evaluated
through the calculation of the determinant of an antisymmetric matrix. The whole investigation necessitates the numerical diagonalization of $2L \times 2L$ matrices with the real time $t$ entering as a parameter in the calculation. Details of the free-fermionic techniques can be found in~\cite{irl}.

In the actual calculation of the magnetization profile we used open chains of length up to $L=1024$ for initially ordered defects, while in the general case the largest size was $L=512$. 

Finite-size values of the defect exponents can be estimated by exploiting
the properties of the expression conjectured for $m_d(t)$ in equation~\eref{m_tL}.
The defect magnetization is calculated in a system of size $L$ at times $t=L/2$ and $t=3L/4$ and 
in a system of size $L/2$ at time $t=L/4$. Forming the ratios~\footnote[1]{Due to the sine, the value of the defect magnetization in equation~\eref{m_tL} should be the same at $t=L/4$ and $t=3L/4$,
however at $L/4$ some oscillating corrections to the leading behavior are often not negligible, see next section.}
\begin{eqnarray}
r(L)&=&m_d(t=L/2,L)/m_d(t=L/4,L/2)\,,\nonumber\\
r'(L)&=&m_d(t=L/2,L)/m_d(t=3L/4,L)\,,
\label{rL}
\end{eqnarray}
the following combinations of exponents are asymptotically obtained
\begin{equation}
\frac{\ln r(L)}{\ln 2}=\alpha(L) \to -x_1+2(x_{12}-x_2)\,,
\label{alpha}
\end{equation}
and
\begin{equation}
\label{alpha'}
\frac{2\ln r'(L)}{\ln 2}=\alpha'(L) \to 2(x_{12}-x_2)\,.
\end{equation}
Note that the relation in equation~\eref{alpha} is a consequence of scaling alone and does not depend 
on the functional form of the time evolution.
In contrast, equation~\eref{alpha'} requires the sine function to be valid.
These exponent combinations have been calculated for sizes up to $L=4096$.

\subsection{Ordered defect in the initial state}
\label{sec:num_fixed}

\begin{figure}[!t]
\begin{center}
\includegraphics[width=9cm,angle=0]{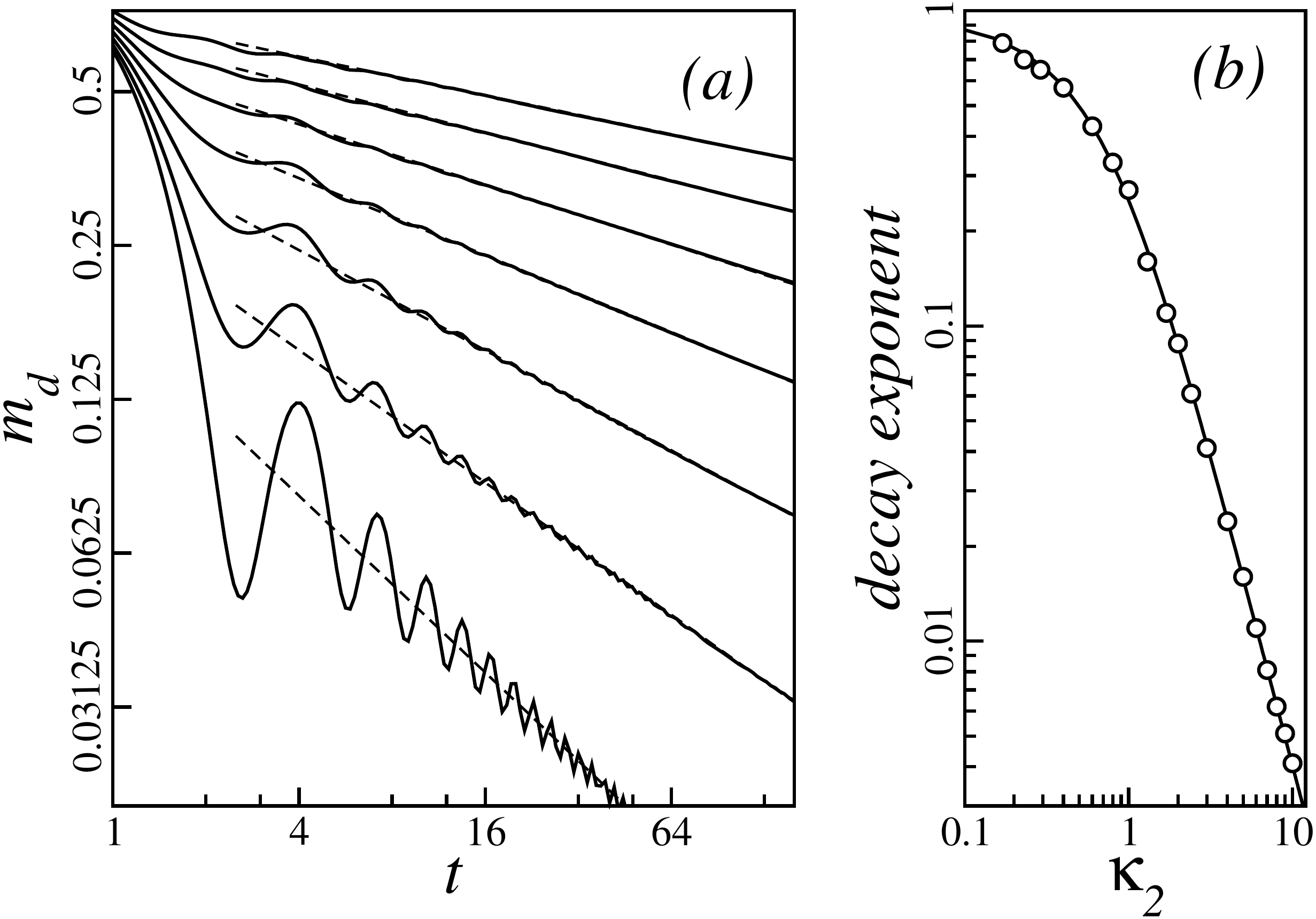}
\end{center}
\vglue -.5cm
\caption{
\label{fig_1} (a) Log--log plot of the relaxation of the defect magnetization in a finite system of length $L=1024$ after a quench from an ordered defect initial state. The values of $\kappa_2$ in the final state vary from $1.6$ to $0.4$, in steps of $0.2$, from top to bottom. The dashed lines indicate the expected decay as given in equation~\eref{m_t}. Note that the amplitude of the decaying initial oscillations is increasing with decreasing $\kappa_2$. (b) Estimated values (circles) of the decay exponents in a system of length $L=512$ for different values of $\kappa_2$ in a log--log scale. The error of the estimate is smaller than the size of the symbols. The line gives the theoretical prediction $2x_2(\kappa_2)$.
}
\end{figure}

When the defect is initially ordered, i.e., when $\kappa_1=\infty$, two of the three exponents are vanishing, $x_1=x_{12}=0$. Thus
the relaxation of the defect magnetization in equation~\eref{m_t} is governed by $x_2$ alone. In the numerical calculations
the ordered initial state is realized with $h_{11}=h_{12}=0$ and $J_1=1$ and the quench is towards a final state with $h_{21}=h_{22}=1$ and $J_2=\kappa_2$. The evolution of the defect magnetization is shown in a log--log plot in figure~\ref{fig_1}(a) for different values of $\kappa_2$.

%%%%%%%%%% FIG 1  %%%%%%%%%%%%%%%%%%%%%%%%%%%%%%%
%%%%%%%%%% FIG 1  %%%%%%%%%%%%%%%%%%%%%%%%%%%%%%%

%%%%%%%%%% FIG 2  %%%%%%%%%%%%%%%%%%%%%%%%%%%%%%%
\begin{figure}[!t]
\begin{center}
\includegraphics[width=9cm,angle=0]{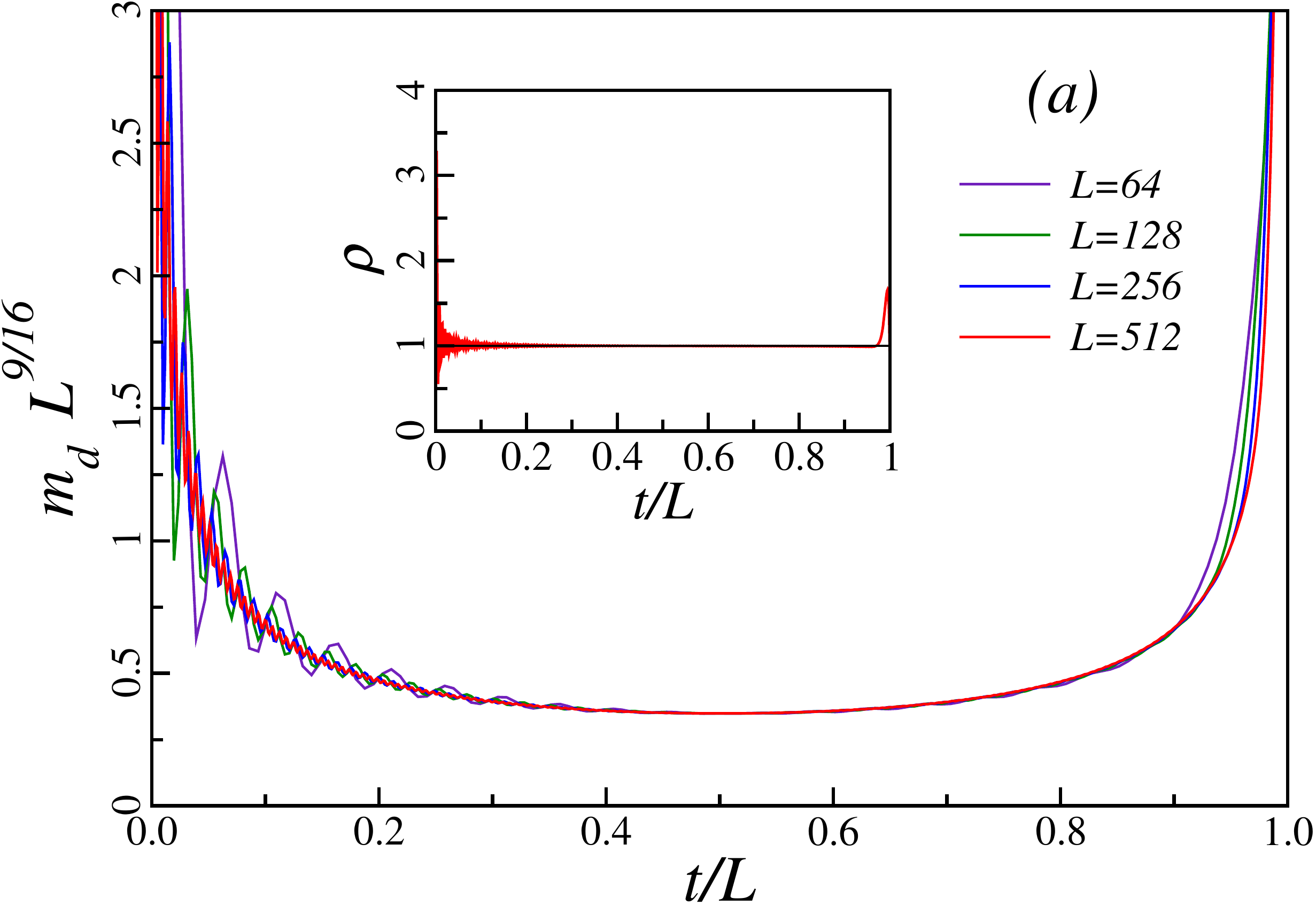}
\vglue 0cm
\includegraphics[width=9cm,angle=0]{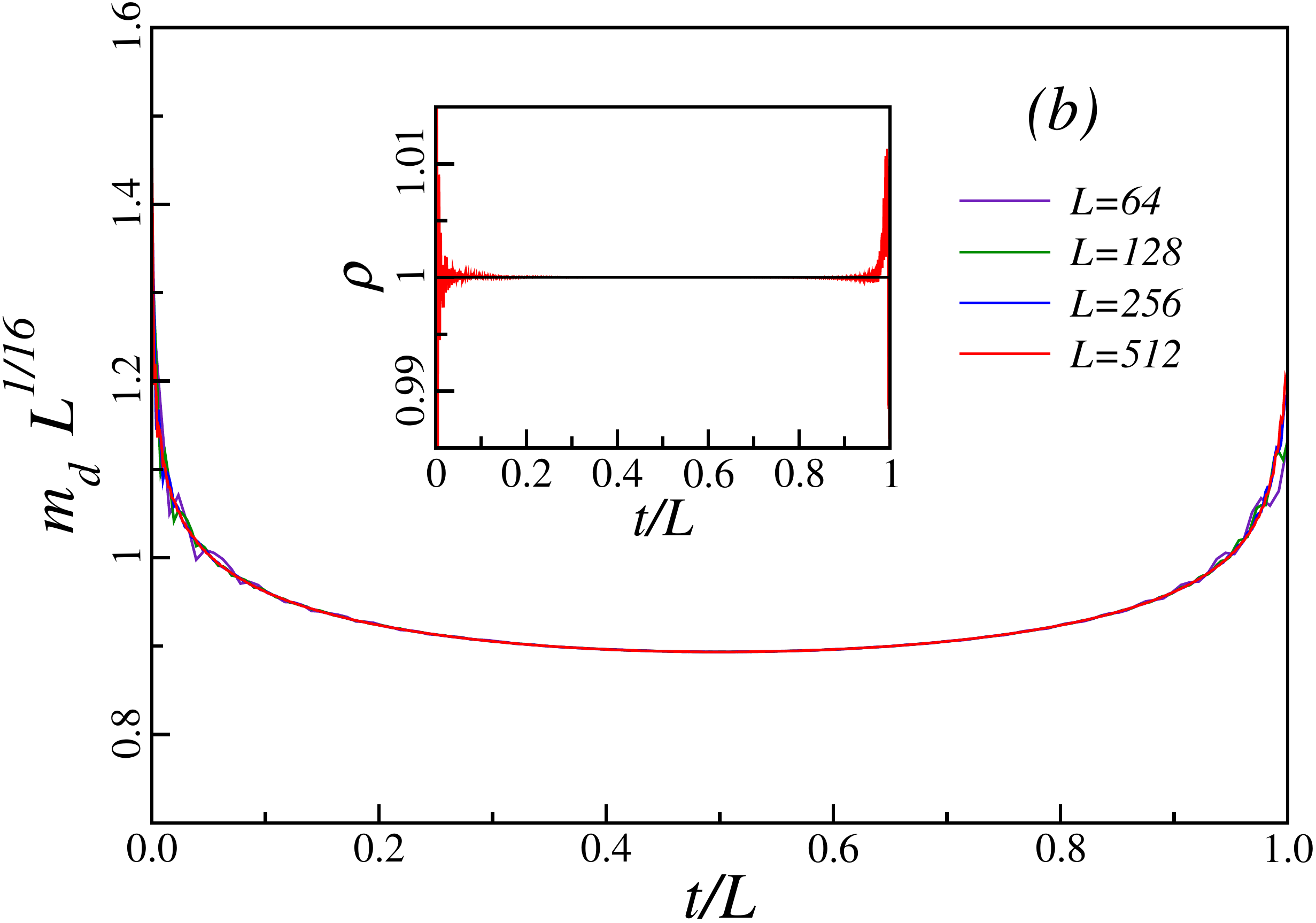}
\end{center}
\vglue -.5cm
\caption{
\label{fig_2} Scaling plot of the defect magnetization for different system sizes
from an initially ordered defect (a) to $\kappa_2=\sqrt{2}-1$ and (b) to $\kappa_2=\sqrt{2}+1$.
In the insets, $\rho$ is the ratio
of the numerical result for $L=512$ to the analytical conjecture in equation~\eref{m_tL}.
}
\end{figure}
%%%%%%%%%% FIG 2  %%%%%%%%%%%%%%%%%%%%%%%%%%%%%%%

When the quench is performed to the homogeneous chain, $\kappa_2=1$ (middle curve in figure~\ref{fig_1}(a)), the decay of the magnetization follows the conformal prediction: $m_d^{(+)}(t) \sim t^{-1/4}$, see equation~\eref{m_+t}. This has been verified before in~\cite{Divakaran_11}.
When the coupling at the defect is increased with respect to the bulk value, $\kappa_2>1$, the decay becomes slower and slower and the corresponding exponent is in good agreement with the value $2x_2$ given by equation~\eref{m_t}. 
For weakened defect couplings $\kappa_2<1$
the relaxation presents a decaying oscillatory modulation around the expected value, the amplitude of which is increasing with
decreasing $\kappa_2$. For sufficiently small $\kappa_2$ and for early times the local magnetization may even change sign. 
In order to deduce a decay exponent in this case too, we have measured the series of minimum [$m_{min}(t_i)$]
and maximum values [$m_{max}(t_i)$] of the oscillations. Their average, $m_{av}(t_i)=[m_{max}(t_i)+m_{min}(t_i)]/2$,
is expected to represent an effective (non-oscillating) decay and this series has been used to analyze
the power-law decay. The estimated values of the decay exponent as a function of $\kappa_2$, shown in figure~\ref{fig_1}(b), are in quite good agreement  with the theoretical prediction of equations~\eref{m_t} and~\eref{x_d}.

\Table{\label{table:1}
Finite-size estimates of the relaxation exponent at the defect as defined in equations~\eref{alpha} and \eref{alpha'}
after a quench from an ordered defect to different values of $\kappa_2$. In the last line the conjectured exact results
are given.}
\br\ms
&\centre{2}{$\kappa_2=\sqrt{2}+1$}&\centre{2}{$\kappa_2=1$}&\centre{2}{$\kappa_2=\sqrt{2}-1$} \\
\ns
&\crule{2}&\crule{2}&\crule{2}\\
L    & $\alpha(L)$ & $\alpha'(L)$ & $\alpha(L)$ & $\alpha'(L)$ & $\alpha(L)$ & $\alpha'(L)$ \\
\mr
128  & -0.06208  & -0.06243   & -0.25305  & -0.25758   & -0.55512  & -0.65624     \\
256  & -0.06235  & -0.06293  & -0.25224  & -0.25308  & -0.57922  &    -0.62445   \\
512  & -0.06266  &  -0.06267 &  -0.25118 &  -0.25072  & -0.60196  &    -0.54546   \\
1024 &  -0.06258 &  -0.06247  & -0.25008  & -0.25056   & -0.54929  &    -0.57194  \\
2048 & -0.06248  &  -0.06254 & -0.25015  &  -0.25027  & -0.56824    &   -0.56046 \\
4096 & -0.06251  &  -0.06250 & -0.25006  &  -0.25013   & -0.56219  &   -0.56107 \\
\mr
Exact & -0.0625  &  -0.0625 & -0.25  & -0.25    & -0.5625  &   -0.5625   \\
\br
\end{tabular}
\end{indented}
\end{table}

We have also studied the time dependence of the oscillation amplitudes, $\Delta m(t_i)=m_{max}(t_i)-m_{min}(t_i)$. The decay is algebraic, $\Delta m(t) \sim t^{-a}$, with an exponent $a=1.6(2)$ which has only a weak
$\kappa_2$ dependence, at least for small values of $\kappa_2$.

To understand the origin of these oscillations let us consider the quench to two uncoupled half chains, i.e., to $\kappa_2=0$.
In this case the problem is reduced to the decay of the surface magnetization starting from a fixed surface spin.
Using the notation of~\cite{irl} (equation~(A.7)) we can write $m_d^{(+)}(t)=P_{1,1}(t)$, the value of which
can be calculated at the critical point by the method outlined in~\cite{igloi_rieger}. In the thermodynamic limit this is given by
\begin{equation}
 m_d^{(+)}(t)=\frac{J_1(2t)}{t} \simeq  \frac{t^{-3/2}}{\sqrt{\pi}} \cos\left( 2t - \frac{3}{4}\pi\right),\quad \kappa_2=0\;.
 \label{m_1+0}
\end{equation}
Here $J_1(x)$ denotes the Bessel function of the first kind for which the asymptotic behavior at large $t$ is indicated. Thus for $\kappa_2=0$ the surface magnetization has just an oscillatory behavior and the corresponding
amplitude decays with an exponent  $a=3/2$. This value agrees within the error of the calculations with the values of $a$ obtained for $\kappa_2>0$.

The numerical results for $\kappa_2>0$ and the exact limiting behavior for $\kappa_2=0$ suggest that the leading
time dependence of $m_d^{(+)}(t)$ is governed by two terms. The dominant asymptotic behavior is given by the power-law 
in equation~\eref{m_t} which is supplemented by an oscillating correction as in equation~\eref{m_1+0}. Thus we expect the following form:
\begin{equation}
  m_d^{(+)}(t) \simeq A(\kappa_2) t^{-2x_2(\kappa_2)}+B(\kappa_2)t^{-a} \cos(2t+\phi)\;.
  \label{osc}
\end{equation}
Here $a \approx 1.5$ and the prefactors, $A(\kappa_2)$ and $B(\kappa_2)$ are even functions of $\kappa_2$ due to symmetry.
For small $\kappa_2$ we have $A(\kappa_2) \sim \kappa_2^2$ and $B(\kappa_2) \simeq 1/\sqrt{\pi}-b \kappa_2^2$.

Next we study the finite-size behavior of the defect magnetization which, according to equation~\eref{m_tL}, has an oscillatory
time dependence of period $L$~\footnote[2]{Note that for $t>L$ the absolute value of the sine has to be taken in equation~\eref{m_tL}.}. In a semi-classical approach this can be explained in terms of ballistically moving quasi-particles, which are reflected at the free boundaries of the chain.
The scaled magnetization $L^{2x_2}m_d^{(+)}$ is plotted as a function of $t/L$, for different values of $L$, for $\kappa_2=\sqrt{2}-1$ in figure~\ref{fig_2}(a) and for $\kappa_2=\sqrt{2}+1$ in figure~\ref{fig_2}(b).
For small values of $t/L$, in particular for small $\kappa_2$, there are strong oscillations in agreement with equation~\eref{osc}. For larger $t/L$, as well as for not too small $\kappa_2$,
the scaled functions present a good data collapse, well described by the conjectured
expression in~\eref{m_tL}. This is illustrated in the insets of figure~\ref{fig_2} in which the ratio $\rho$ of the numerical results
for $L=512$ and the analytical conjecture in~\eref{m_tL} is shown as a function of $t/L$. For the latter the prefactor is fixed in order to have a ratio $\rho=1$ at
$t=L/2$.

Finite-size estimates for the defect exponent $x_2$ has been obtained using the relations in equations~\eref{alpha} and
\eref{alpha'}. The results collected in table \ref{table:1} converge rather well to the
expected exact values, which confirms also that the sinusoidal form of the profile in equation~\eref{m_tL} is probably exact. Again, for
the smallest value of $\kappa_2$, $\sqrt{2}-1$, due to the oscillations the data are non-monotonic in $L$.

%%%%%%%%%% FIG 3  %%%%%%%%%%%%%%%%%%%%%%%%%%%%%%%
\begin{figure}[!t]
\begin{center}
\includegraphics[width=9cm,angle=0]{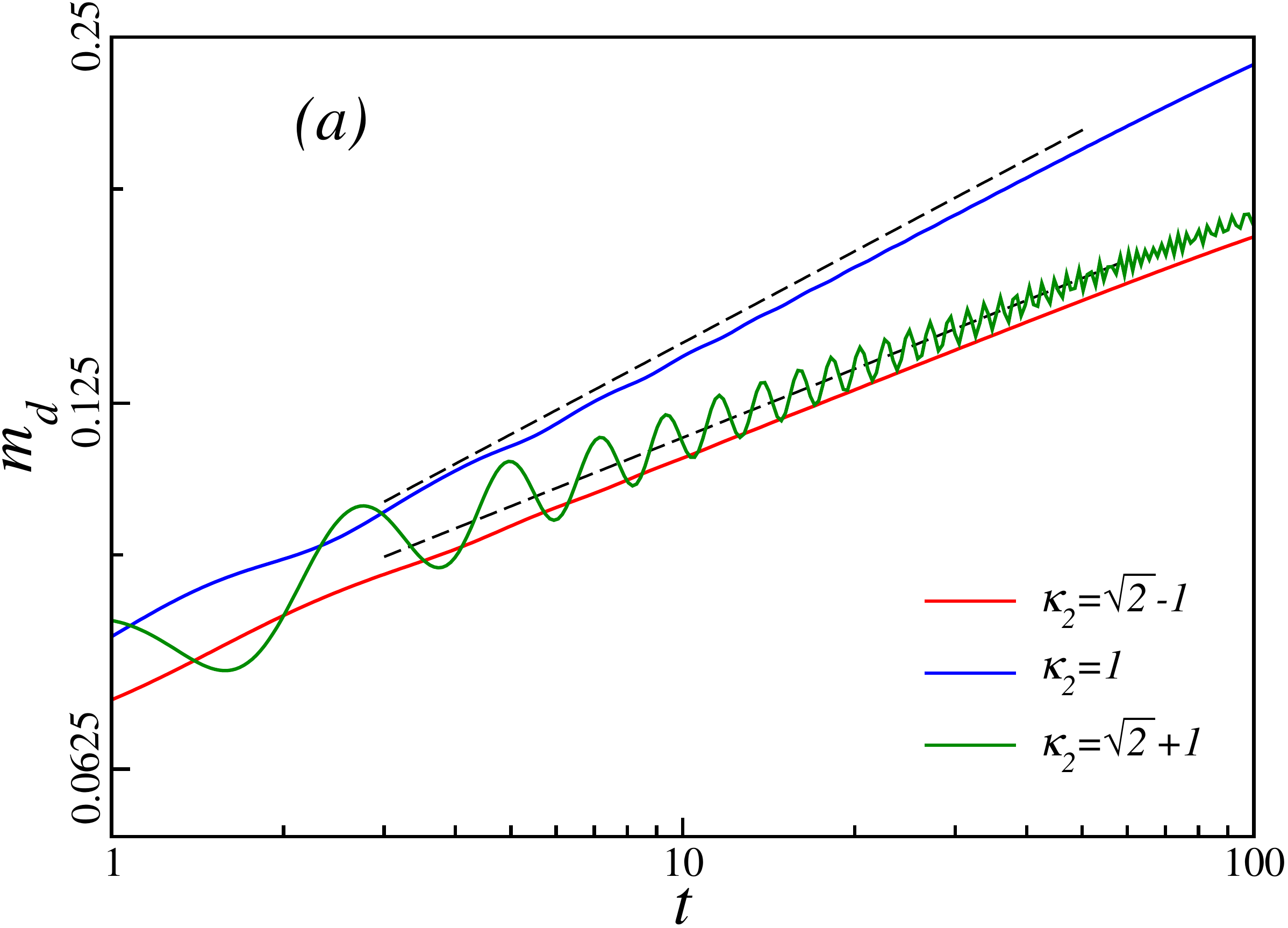}
\vglue 0cm
\includegraphics[width=9cm,angle=0]{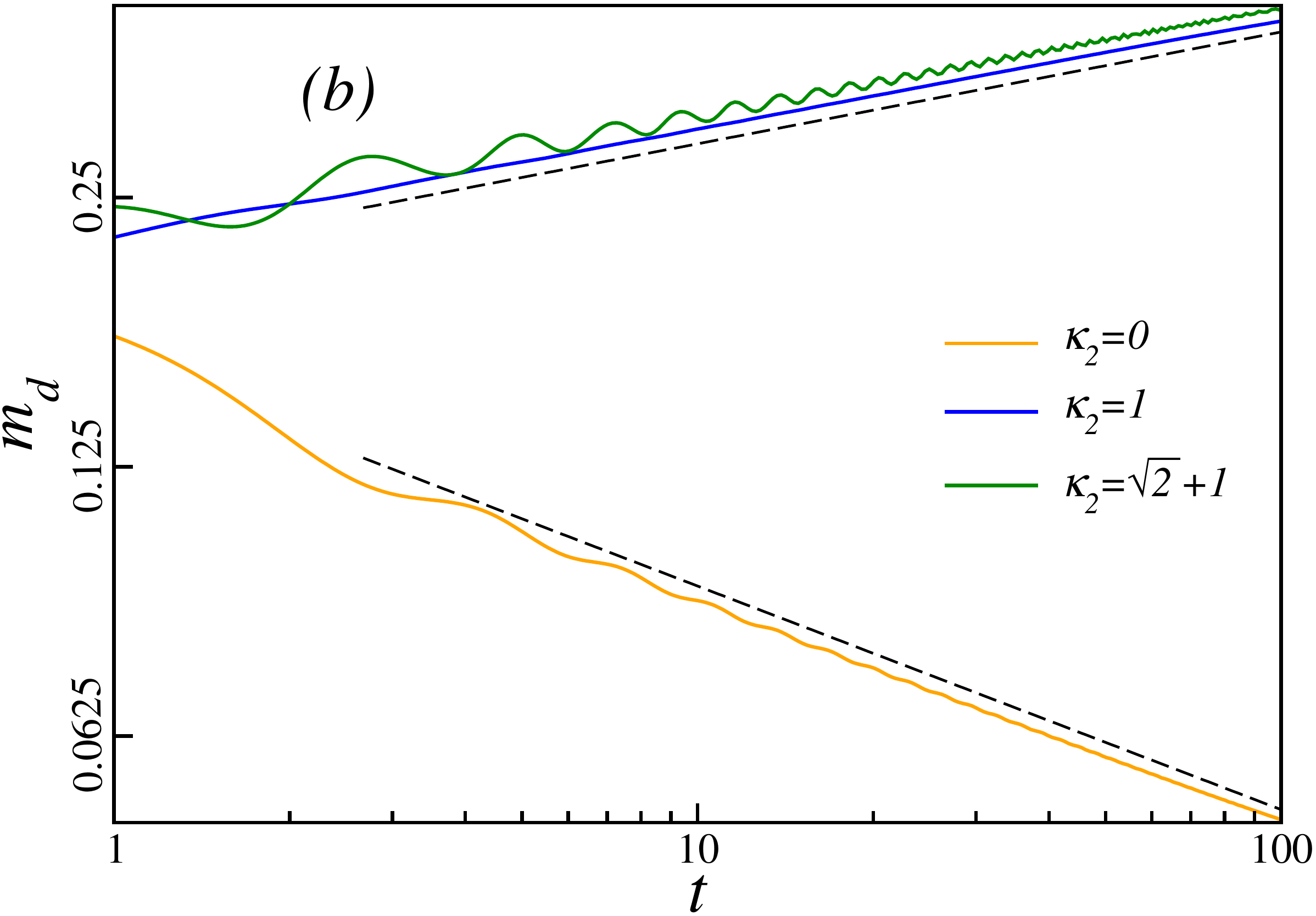}
\end{center}
\vglue -.5cm
\caption{
\label{fig_3} Log--log plot of the time dependence of the defect magnetization in a chain of length $L=512$ 
after a local quench, when the strength of the defect changes from (a) $\kappa_1=0$ and (b) $\kappa_1=\sqrt{2}-1$ 
to different values of $\kappa_2$. The dashed lines indicate the analytical result
in equation~\eref{m_t}. Note that in both figures there are two quenches with the same decay exponent.
}
\end{figure}
%%%%%%%%%% FIG 4  %%%%%%%%%%%%%%%%%%%%%%%%%%%%%%%
\begin{figure}[!t]
\begin{center}
\includegraphics[width=9cm,angle=0]{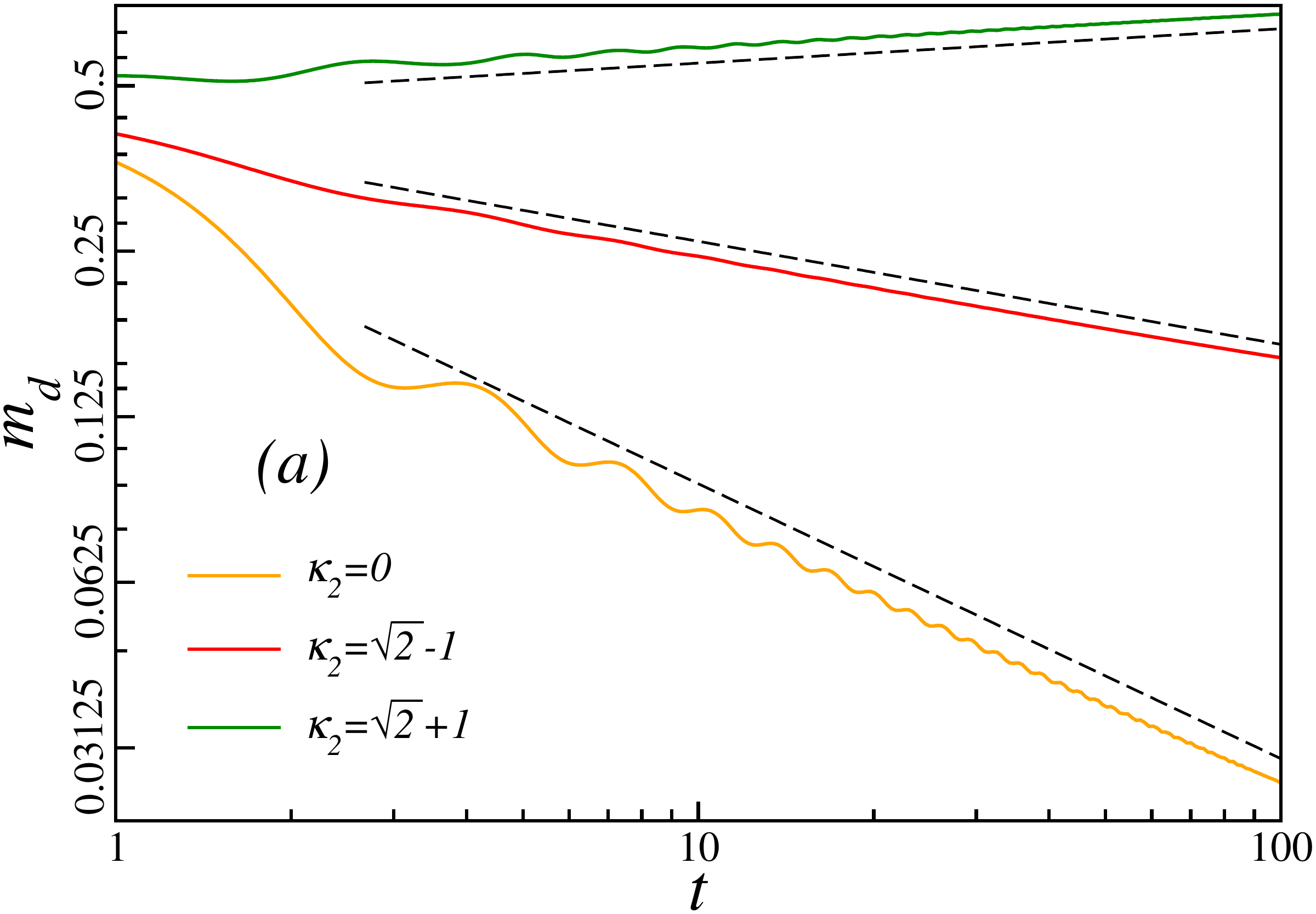}
\vglue 0cm
\includegraphics[width=9cm,angle=0]{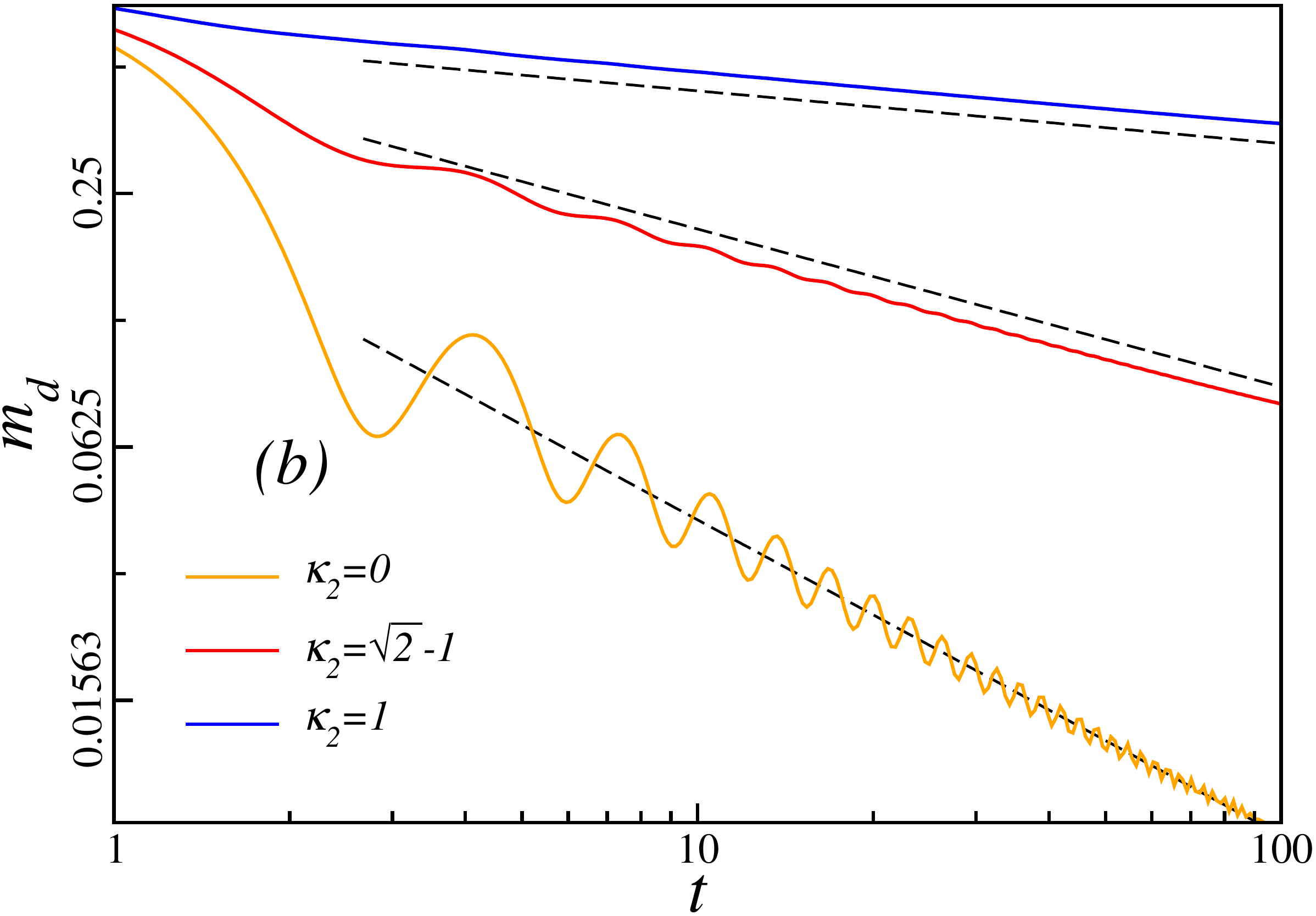}
\end{center}
\vglue -.5cm
\caption{
\label{fig_4} As in figure~\ref{fig_3}, for (a) $\kappa_1=1$ and (b) $\kappa_1=\sqrt{2}+1$.
}
\end{figure}
%%%%%%%%%% FIG 4  %%%%%%%%%%%%%%%%%%%%%%%%%%%%%%%

%%%%%%%%%% FIG 5  %%%%%%%%%%%%%%%%%%%%%%%%%%%%%%%
\begin{figure}[!t]
\begin{center}
\includegraphics[width=9cm,angle=0]{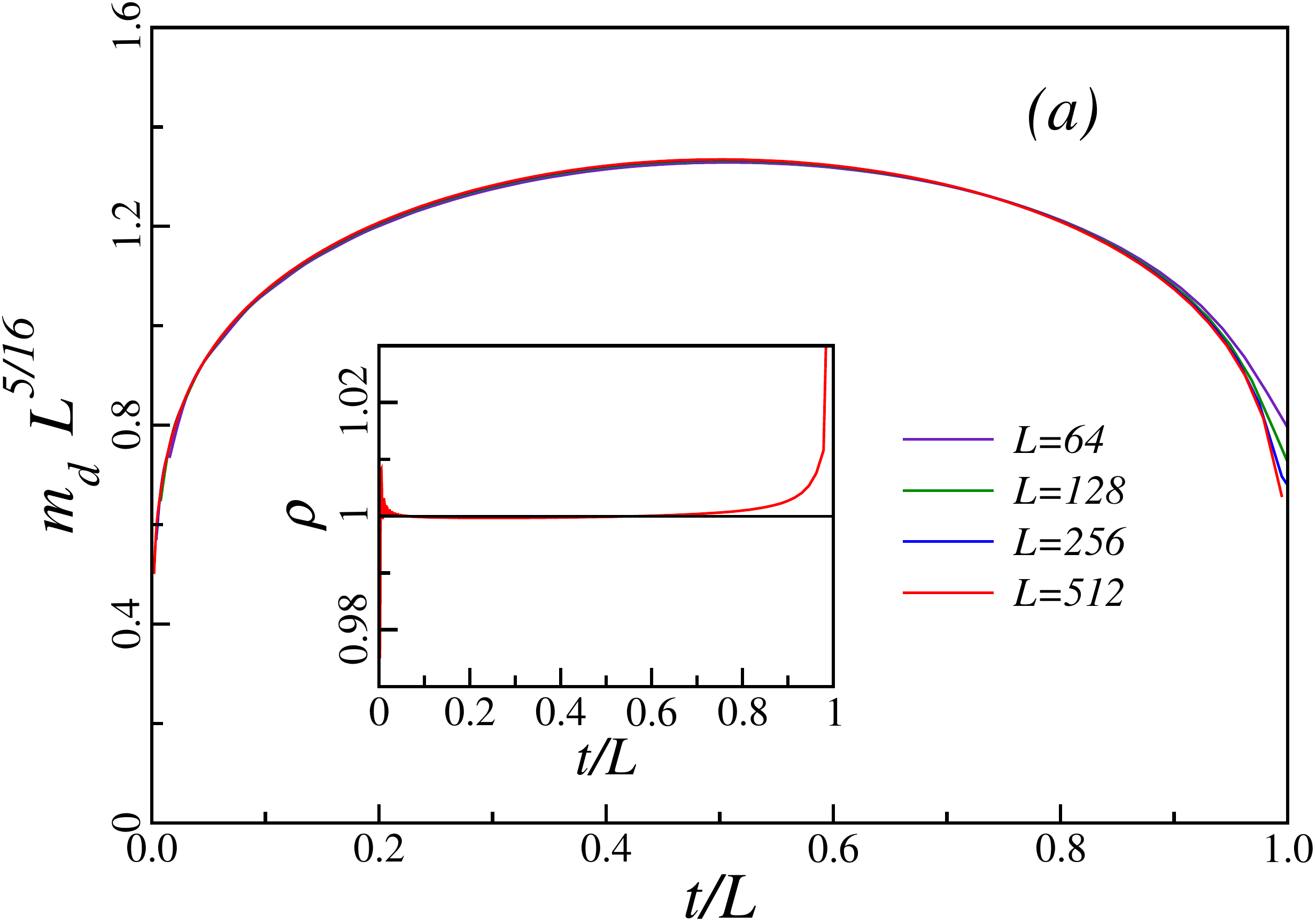}
\vglue 0.cm
\includegraphics[width=9cm,angle=0]{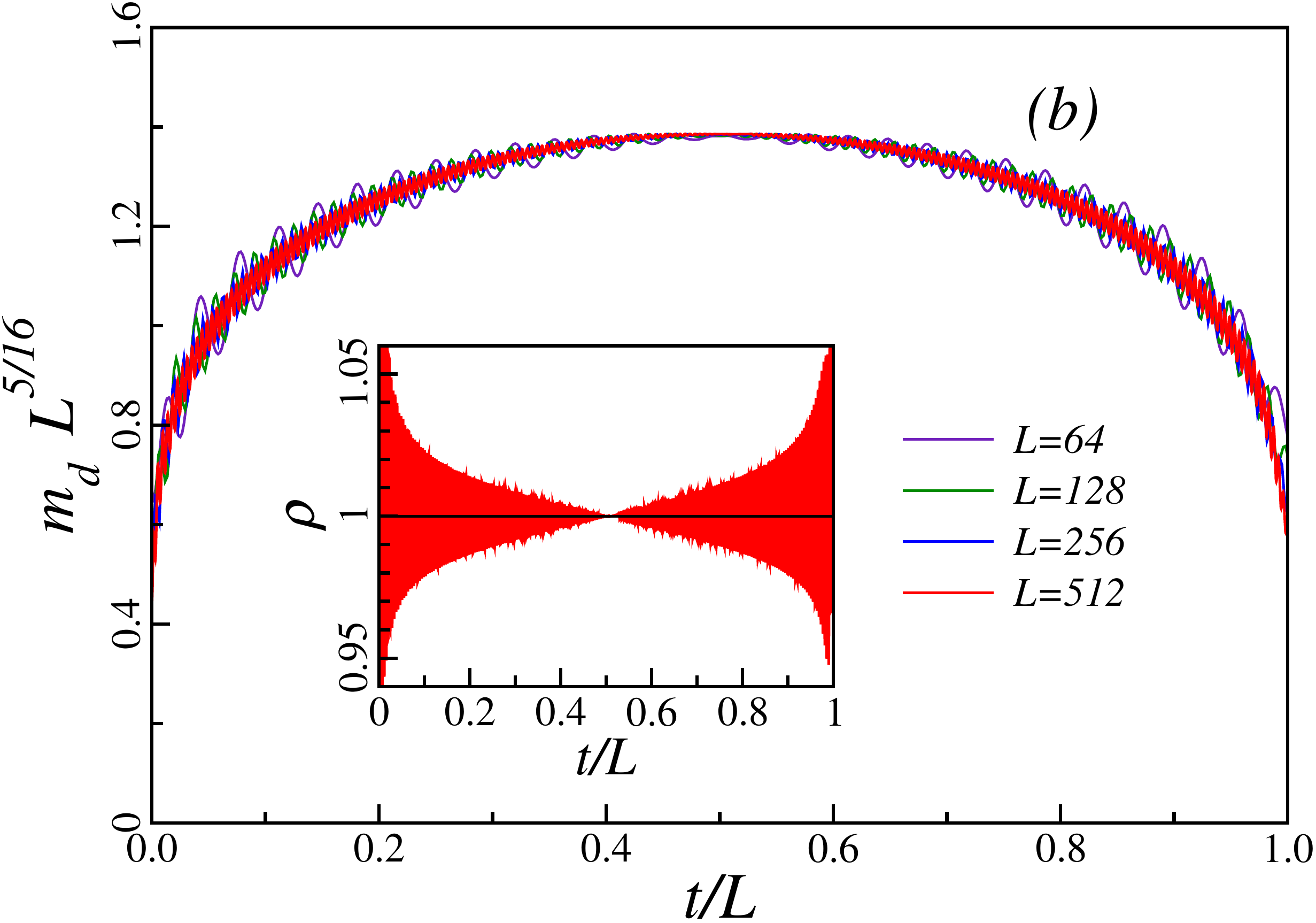}
\end{center}
\vglue -.5cm
\caption{
\label{fig_5} Scaling plot of the defect magnetization for different system sizes for a quench 
from an initial
state with $\kappa_1=0$ to a final state with (a) $\kappa_2=\sqrt{2}-1$ and (b) $\kappa_2=\sqrt{2}+1$. The inset gives the ratio $\rho$ 
of the numerical result for $L=512$ to the analytical conjecture in equation~\eref{m_tL}.
}
\end{figure}
%%%%%%%%%% FIG 5  %%%%%%%%%%%%%%%%%%%%%%%%%%%%%%%

\subsection{Non-ordered defect in the initial state}
\label{sec:num_defect}

For initially non-ordered defects we use the parameters $h_{11}=h_{12}=1$
and $J_1=\kappa_1$ and the quench is performed to $h_{21}=h_{22}=1$ and $J_2=\kappa_2$. The values of $\kappa_1$ and $\kappa_2$ are taken from the set $\{0,\tan(\pi/8)=\sqrt{2}-1,1,1/\tan(\pi/8)=\sqrt{2}+1\}$. The defect exponents
are then rational numbers $\{1/2,9/32,1/8,1/32\}$, respectively, and the same is true for the
composite defect exponents. The time dependence of the defect magnetization is shown in log--log plots 
for different values of the initial and final defect couplings with $\kappa_1=0$ and $\sqrt{2}-1$ in figure~\ref{fig_3}, 
$\kappa_1=1$ and $\sqrt{2}+1$ in figure~\ref{fig_4}. The values of $\kappa_2$ are the remaining ones in the set given above.
In agreement with the scaling results the curves have a linear starting behavior and the
slope is well described by the analytical expression in~\eref{m_t}. The short-time behavior is more or less oscillating, depending on the relative strength of the defect, before and after the quench.

The finite-size behavior of the defect magnetization have been also studied, in order to check
the scaling prediction and the functional form given in equation~\eref{m_tL}. The scaled defect magnetization $m_d(t,L)L^{x_1-2(x_{12}-x_2)}$ 
is shown in figure~\ref{fig_5} for $\kappa_1=0$ and for two values of $\kappa_2$. An excellent data collapse is obtained for the smaller value $\kappa_2=\sqrt{2}-1$ (figure~\ref{fig_5}(a)). For $\kappa_2=\sqrt{2}+1$
(figure~\ref{fig_5}(b)) the collapse is perturbed by the oscillations, which are stronger at smaller sizes. In both cases the overall trend confirms the conjectured result in equation~\eref{m_tL}. This is well illustrated in the insets, where the ratio $\rho$ of the numerical results
for $L=512$ to the analytical conjecture in~\eref{m_tL} is shown. Once more the amplitude of the latter is chosen so that $\rho=1$ at
$t=L/2$.

\Table{\label{table:2}
Finite-size estimates of the relaxation exponent at the defect as defined in equations~\eref{alpha} and \eref{alpha'}
after a quench from $\kappa_1=\sqrt{2}-1$ to different values of $\kappa_2$. The conjectured exact results
are given in the last line.}
\br\ms
&\centre{2}{$\kappa_2=0$}&\centre{2}{$\kappa_2=1$}&\centre{2}{$\kappa_2=\sqrt{2}+1$} \\
\ns
&\crule{2}&\crule{2}&\crule{2}\\
L    & $\alpha(L)$ & $\alpha'(L)$ & $\alpha(L)$ & $\alpha'(L)$ & $\alpha(L)$ & $\alpha'(L)$ \\
\mr
128  & -0.52609  &  -0.23745  & -0.15460  & -0.12094   & -0.15554  & 0.13977     \\
256  & -0.52870  &  -0.24379 & -0.15547  &  -0.12299  & -0.15528  &   0.12397    \\
512  &  -0.52970 &  -0.24679 & -0.15586  &   -0.12399 & -0.15584  &   0.12133    \\
1024 &  -0.53061 &  -0.24839  & -0.15606  &  -0.12449  & -0.15607  &   0.12086   \\
2048 & -0.53093  &  -0.24923 & -0.15616  &  -0.12474  &  -0.15617   & 0.12476   \\
4096 & -0.53109  &  -0.24962 &  -0.15620 &  -0.12487   & -0.15620  &  0.12446 \\
\mr
Exact & -0.53125  &  -0.25 & -0.15625  &  -0.125   & -0.15625  &  0.125    \\
\br
\end{tabular}
\end{indented}
\end{table}

%%%%%%%%%% FIG 6  %%%%%%%%%%%%%%%%%%%%%%%%%%%%%%%
\begin{figure}[!t]
\begin{center}
\includegraphics[width=9cm,angle=0]{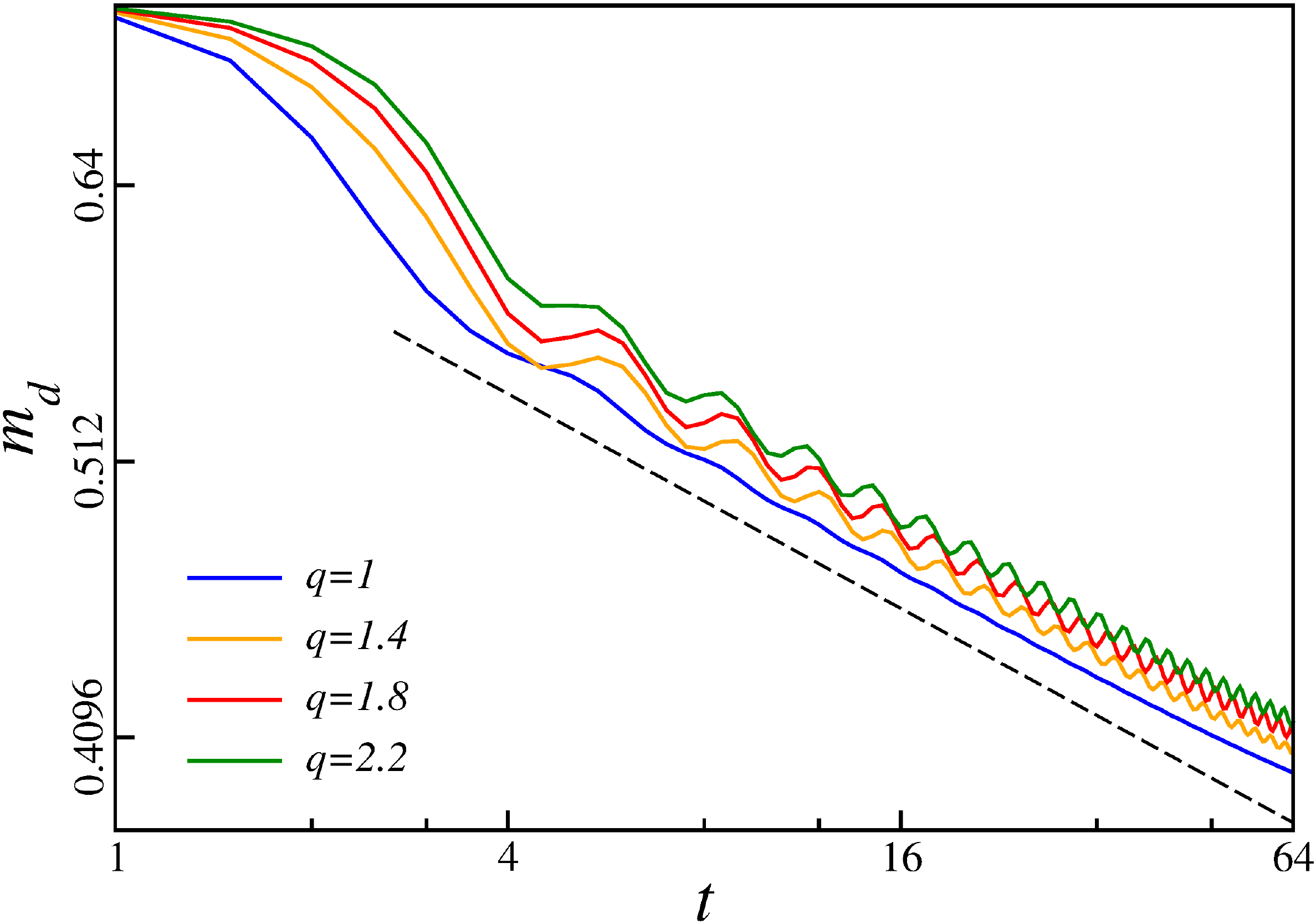}
\end{center}
\vglue -.5cm
\caption{
\label{fig_6} Log--log plot of the time dependence of the defect magnetization in a chain of length $L=256$ with a more complex defect structure after the quench. In the initial state $J_1=1+\sqrt{2}$, $h_{11}=h_{12}=1$ thus
$\kappa_1=1+\sqrt{2}$ and $x_1=1/32$. In the final state $J_2=1$, $h_{21}=\sqrt{q}$, $h_{22}=1/\sqrt{q}$ thus $\kappa_2(q)=1$, $x_2=1/8$ and $x_{12}=1/16$. The slopes for the different values of $q$ are in good agreement with the expected one, $2(x_{12}-x_2)=-1/8$ (dashed line).}
\end{figure}
%%%%%%%%%% FIG 5  %%%%%%%%%%%%%%%%%%%%%%%%%%%%%%%

Quantitative estimates of the defect exponents have been obtained using the relations given in equations~\eref{alpha} 
and~\eref{alpha'} for finite-size results on chains with sizes up to $L=4096$. The values of the effective exponents 
for $\kappa_1=\sqrt{2}-1$ and different values of $\kappa_2$ are shown in table \ref{table:2}.
For each combination of $\kappa_1$ and $\kappa_2$ the effective exponents are found to converge to the conjectured values.
The finite-size estimate for $\alpha(L)$ at the largest size agrees with the conjectured value up to four or five digits
except when $\kappa_2$ is small, the defect magnetization showing then strong oscillations (see equation~\eref{osc}).
Very good, although somewhat less accurate estimates are found for the exponent $\alpha'(L)$, too. This confirms that
the conjecture about the finite-size scaling form in equation~\eref{m_tL} is most probably exact.

\section{Discussion}
We have studied the evolution of the local magnetization in the TIC after a quench, when
parameters at a defect are suddenly modified. At short time,  the defect magnetization displays a power-law
behavior which is closely related to the local static critical behavior at a composite line defect in 
the $2d$ classical Ising model, which corresponds to the imaginary time version of our problem. The composite defect exponents have been exactly calculated making use of conformal invariance. Since the perturbation is truly marginal the local magnetic exponents are continuously varying with the parameters of the composite defect, i.e., their values before and after the quench in real time. 

In finite chains the defect magnetization is
a periodic function of time and we have conjectured its asymptotic functional form. The analytical expressions, both
for $t \ll L$, as well as for $t/L={\cal O}(1)$, have been confronted to the results of large-scale numerical calculations and an excellent agreement has been found.
Therefore we expect that equations~\eref{m_t} and~\eref{m_tL} are a consequence of conformal invariance and can be derived rigorously.

The defect exponents $x_1$ and $x_2$, as well as the composite defect exponent $x_{12}$, are functions of the defect parameters
$\kappa_1$ and $\kappa_2$, as defined in equation~\eref{kappa_ii}. We have checked numerically (see figure 6 for an illustration) that details of the local
defect structure (asymmetry in the transverse fields, etc.) are indeed irrelevant and that only the values of $\kappa_i$
matter in this respect. The same property is expected to hold for the critical entanglement entropy across a defect.

As mentioned above, the non-universal local critical behavior found in our study is a specificity of the TIC and is related to the fact that the perturbation caused by a localized defect is truly marginal for this system.
In other models such a thermal perturbation may be irrelevant or relevant, depending on the value of the correlation length exponent $\nu$ in the unperturbed system~\cite{igloi93}. For $\nu >  1$,
as for the $Q$-state Potts model with $Q<2$ (such as uncorrelated percolation), the defect is an irrelevant perturbation 
and the local critical behavior at the defect is the same as in the bulk of the system, thus $x_1=x_2=x_{12}=x_m$ for a weak local perturbation. In such models
a local quench has no effect on the non-equilibrium
properties of the system. For $\nu <1$, such as for a quantum Potts chain with $2<Q\leq 4$, the defect is a relevant perturbation
and the local critical behavior is different for $\kappa<1$ and for $\kappa>1$.
For weakened local couplings, $\kappa<1$, the defect is renormalized to a cut and the defect exponent is the same as at a
free surface, $x_i=x_{ms}$. In contrast, for local couplings stronger than the bulk ones, $\kappa>1$, the defect renormalizes to an ordered one,
consequently $x_i=0$. Then a local quench is efficient only when the fixed points, before and after the quench, are different.
If $\kappa_1>1$ and $\kappa_2<1$, then
$x_2=x_{ms}$ and $x_1=x_{12}=0$~\footnote[3]{Note that, due to local order, the composite-defect exponent $x_{12}$ is always expected to vanish when $x_1=0$.}, and from equation~\eref{m_t} we see that the decay is given by $m_d(t)\sim t^{-2x_{ms}}$.

\ack
This work has been supported by the Hungarian National Research Fund under grant
No OTKA K75324, K77629 and K109577. FI thanks Uma Divakaran
and Heiko Rieger for previous cooperation in the project and the Groupe de Physique Statistique, Institut Jean Lamour, Universit\'e de Lorraine
for hospitality during the starting period of this work; LT thanks Malte Henkel and J\'er\^ome Dubail for useful discussions.

\appendix
\section*{Appendix: Composite defect exponents}
\label{sec:A1}
\setcounter{section}{1}

\begin{figure}[!t]
\vglue 5mm
\begin{center} 
\includegraphics[width=9cm,angle=0]{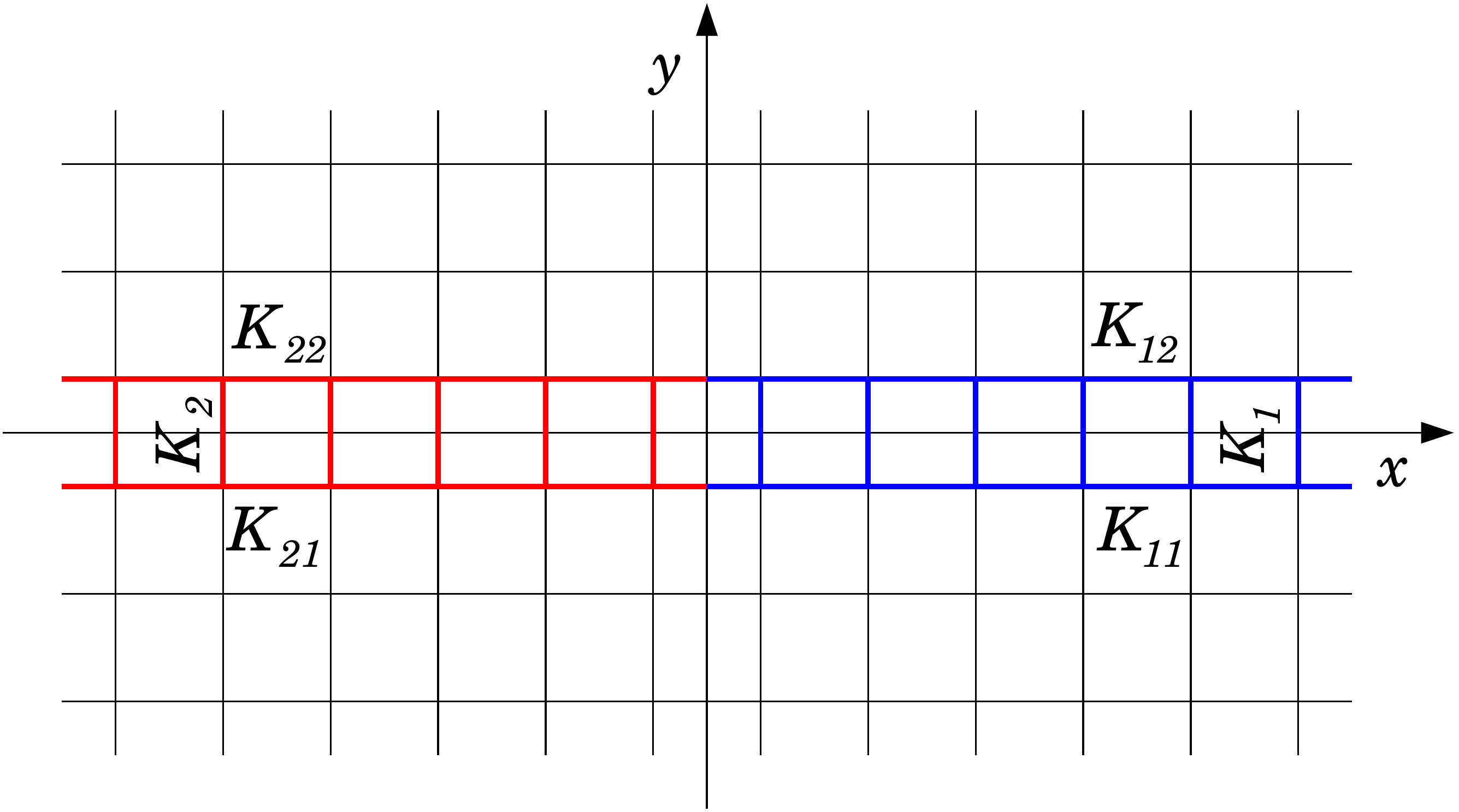}
\end{center}
\vglue-5mm
\caption{Composite line defect in the critical two-dimensional square lattice Ising model. The composite defect is made of two half-lines of perturbed couplings on a ladder.} 
\label{figA1}
\end{figure}

Let us consider the critical Ising model on the square lattice with a line defect as shown in figure~\ref{figA1}. 
The composite defect results from the junction of two semi-infinite line defects, indexed 1 and 2, with different 
horizontal ($K_{ij}$) and vertical ($K_i$) perturbed couplings ($i,j=1,2$).
Since the scaling dimension of the bulk energy density, $x_e=1$, is the same as the dimension of the line defect, the perturbation is marginal and varying local magnetic exponents are expected as for the infinite line defect~\cite{bariev79,mccoy80}. 

In the off-critical system the local behavior of the magnetization, at a distance from the defect smaller than the bulk correlation length $\xi$, is governed by three different exponents. In the central region the local magnetization exponent $x_{12}$ is influenced by the two parts of the composite defect. Outside this region, at a distance larger than $\xi$ from the junction, the local magnetization exponents, $x_1$ and $x_2$, are the same as for infinite line defects.

%%%%%%%%%%% FIG A1  %%%%%%%%%%%%%%%%%%%%%%%%%%%%%%%

%%%%%%%%%%% FIG A2  %%%%%%%%%%%%%%%%%%%%%%%%%%%%%%%

\begin{figure}[!t]
%\vglue 5mm
\begin{center} 
\includegraphics[width=9cm,angle=0]{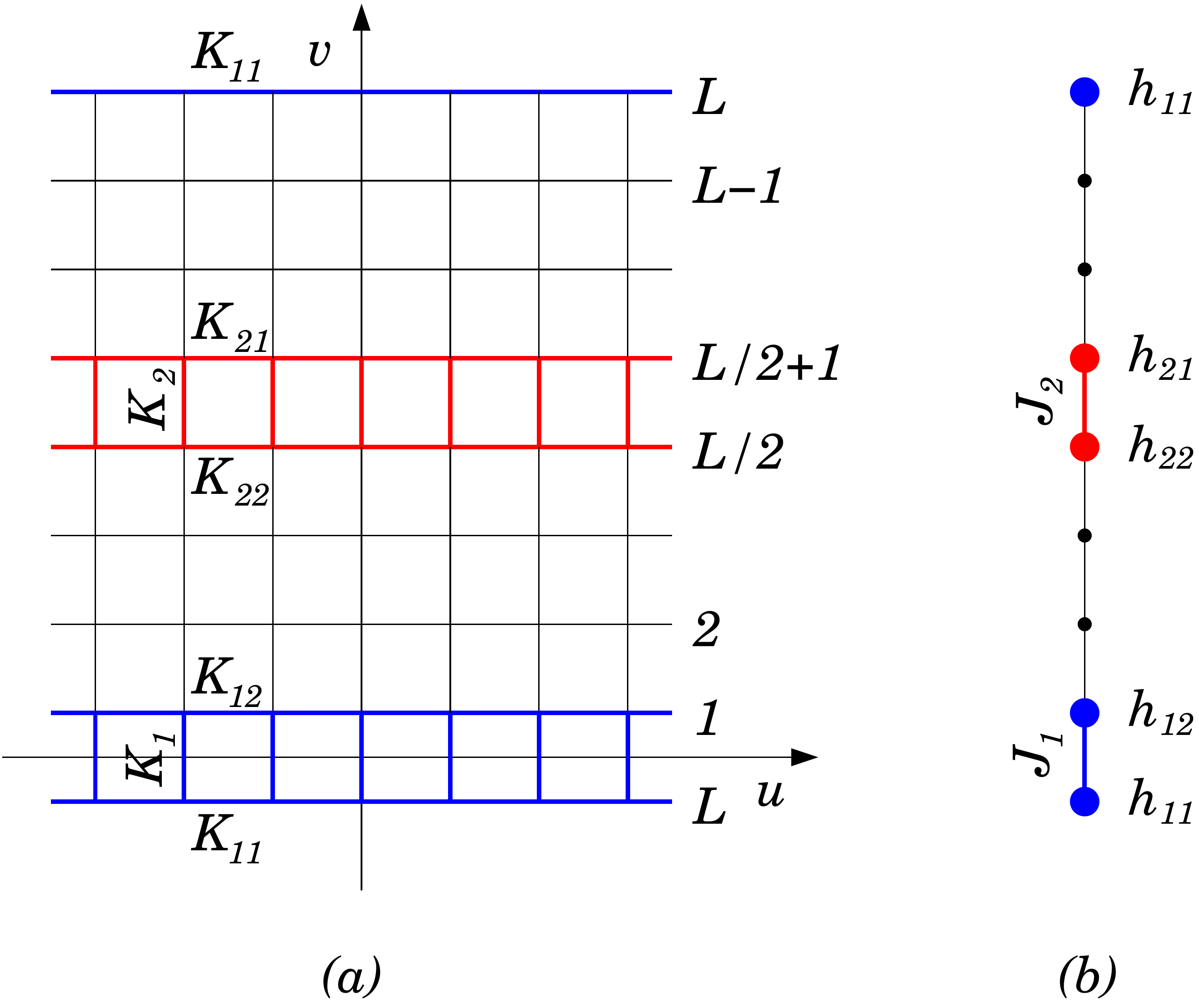}
\end{center}
\vglue-5mm
\caption{(a) Under the conformal transformation $w=(L/2\pi)\ln z$ the full plane with a composite line defect becomes an infinite cylinder with circumference $L$ and two infinite equidistant line defects, parallel to the cylinder axis. (b) In the extreme anisotropic limit, the column-to-column transfer operator can be expressed as the exponential of the Hamiltonian of a TIC, up to a rescaling factor.} 
\label{figA2}
\end{figure}

%%%%%%%%%%%%%%%%%%%%%%%%%%%%%%%%%%%%%%%%%%%%%%

The composite defect exponent $x_{12}$ can be obtained using conformal methods and finite-size scaling~[65]--[70].
%\cite{turban85,guimaraes86,henkel87a,henkel87b,henkel88,henkel89}.  
In a first step, the infinite critical system of figure~\ref{figA1}, with a single composite line defect along the $x$-axis, is transformed into a cylinder with two equidistant line defects, 1 and 2, parallel to the cylinder axis, through the conformal transformation $w=(L/2\pi)\ln z$ where $z=x+iy$ and $w=u+iv$. The transformed system, shown in figure~\ref{figA2}(a), is infinite along the $u$-axis and periodic with size $L$, even, in the transverse direction. In the cylinder geometry the gap-exponent relation~\cite{cardy84a} can be used to extract the composite defect exponents.

Although the column-to-column transfer matrix can in principle be diagonalized for arbitrary couplings, a simplification occurs in the strongly anisotropic (Hamiltonian) limit~\cite{fradkin78,kogut79} where the couplings in the longitudinal direction ($K_u$ in the bulk and $K_{ij}$ on the line defects) are strong while the couplings in the transverse direction ($K_v$ in the bulk and $K_i$ on the line defects) are weak. For a critical bulk in the extreme anisotropic limit, corresponding to a continuous imaginary time along the $u$-axis, the ratio $K_{vc}/K_{uc}^\ast\to 1$ whereas on the line defects 
$K_i/K_{uc}^\ast\to J_i$ and $K_{ij}^\ast/K_{uc}^\ast\to h_{ij}$. 
Then the transfer operator takes the form ${\cal T}=\exp(-2K_{uc}^\ast{\cal H})$ where ${\cal H}$ is the Hamiltonian of a TIC~\cite{pfeuty70} (see figure~\ref{figA2}(b))
\begin{eqnarray}
\fl{\cal H}=-\frac{1}{2}\left[\,\sum_{n=1}^L\sigma_n^x \sigma_{n+1}^x+(J_1-1)\,\sigma_L^x \sigma_1^x+(J_2-1)\,\sigma_{L/2}^x \sigma_{L/2+1}^x\right.\nonumber\\
\fl\ \ \ \ \ \ \ \  +\left.\sum_{n=1}^L\sigma_n^z+(h_{11}-1)\,\sigma_L^z+(h_{12}-1)\,\sigma_1^z
+(h_{22}-1)\,\sigma_{L/2}^z+(h_{21}-1)\,\sigma_{L/2+1}^z\right].
\label{eA1}
\end{eqnarray}

Expressing the Pauli spin operators, $\sigma_n^x$ and $\sigma_n^z$, in terms of fermion operators via the Jordan-Wigner transformation~\cite{jordan28} a quadratic form in fermions is obtained which, due to the periodic boundary conditions, involves an operator ${\cal P}=(-1)^Q$, associated with the bond $(L,1)$ and commuting with ${\cal H}$, with eigenvalues $+1$ ($-1$) corresponding to $Q=0$ ($Q=1$) when the number of fermions is even (odd). In each subspace ${\cal H}(Q)$ is diagonalized by a Bogoljubov transformation~\cite{schultz64,pfeuty70} and takes the form
${\cal H}=\sum _k \epsilon_k\,\left(\eta_k^\dagger\eta_k-\frac{1}{2}\right)$ in terms of the new fermion operators $\eta_k$ and $\eta_k^\dagger$.

The excitation energies $\epsilon_k$ squared are the $Q$-dependent eigenvalues of the $L\times L$ matrix equation
$ ({\textbf M}-\epsilon_k^2)\,{\rm\bf\Phi}_k=0$ with line $n$ ($1<n<L$) given by:
\be\fl
h(n\!-\!1)J(n\!-\!1)\Phi_k(n\!-\!1)+[h^2(n)\!+\!J^2(n\!-\!1)\!-\!\epsilon_k^2]\Phi_k(n)\!+\!h(n)J(n)\Phi_k(n\!+\!1)\!=\!0.
\label{eA2}
\ee
In lines 1 and $L$ one has to replace $J_1$ by $(-1)^{Q+1}J_1$. $h(n)$ is the transverse field at site $n$ and $J(n)$ the coupling between sites $n$ and $n+1$ as shown in figure~\ref{figA2}(b).

With $\epsilon_k^2=4\sin^2(k/2)$ and using the Ansatz 
\begin{eqnarray}
\fl\Phi_k(1)=-A\,,\quad \Phi_k(n)=(-1)^n\left(B\,e^{ikn}+C\,e^{-ikn}\right)\quad (n=2,L/2)\,,\nonumber\\
\fl\Phi_k(L/2\!+\!1)\!=(-1)^{L/2+1}D,\quad \Phi_k(n)=(-1)^n\left(E\,e^{ikn}\!+\!F\,e^{-ikn}\right)\quad (n=L/2\!+\!1,L),
\label{eA3}
\end{eqnarray}
equation~\eref{eA1} becomes a linear system of six equations for the amplitudes $A$ to $F$. The characteristic equation gives the allowed values of $k$ in each $Q$ sector. 

%%%%%%%%%%% FIG A3  %%%%%%%%%%%%%%%%%%%%%%%%%%%%%%%

\begin{figure}[!t]
%\vglue 5mm
\begin{center}
\includegraphics[width=8cm,angle=0]{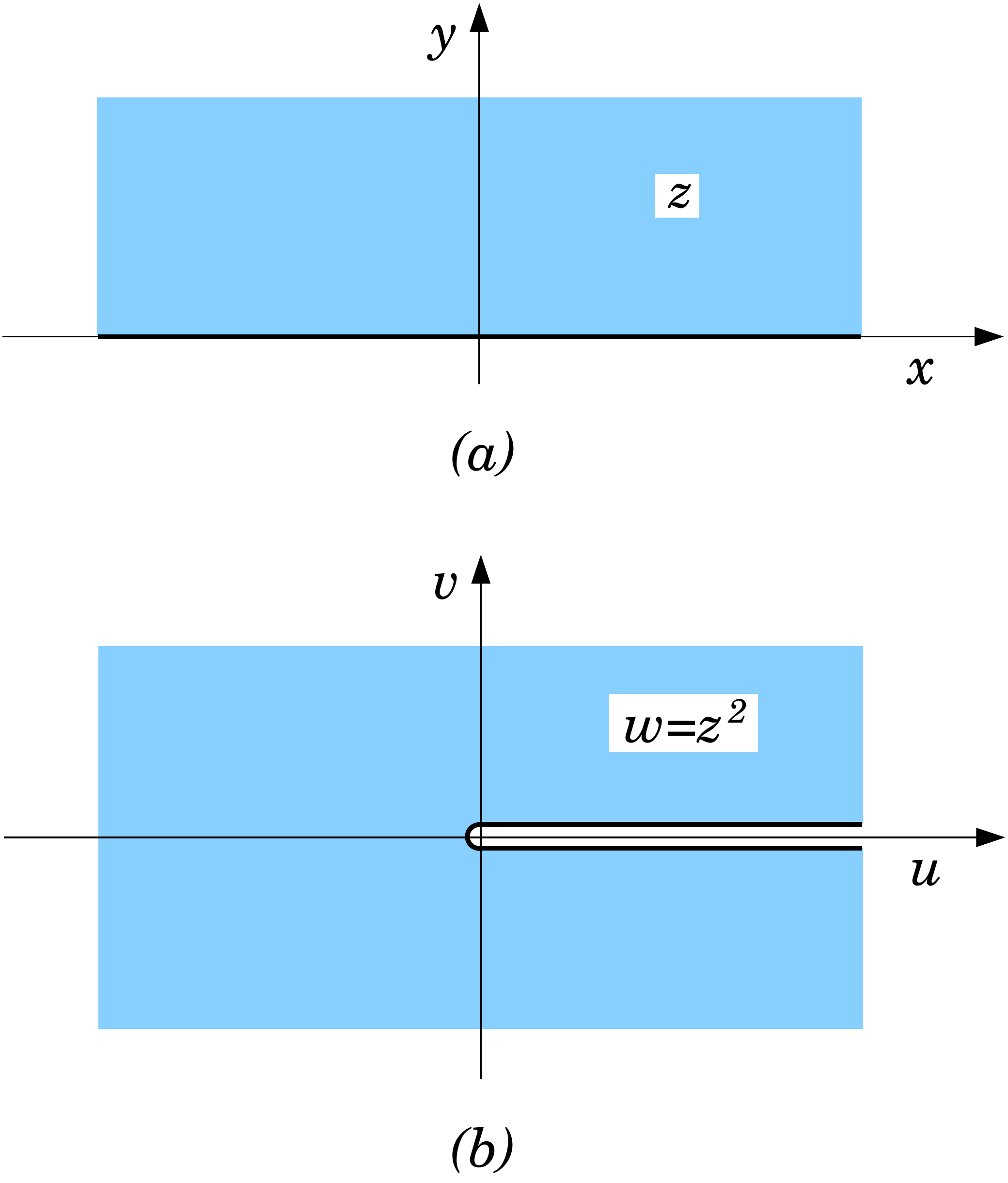}
\end{center}
\vglue-5mm
\caption{The conformal transformation $w=z^2$ maps the upper half-plane onto the full plane with a cut along the positive real axis.} 
\label{figA3}
\end{figure}

%%%%%%%%%%%%%%%%%%%%%%%%%%%%%%%%%%%%%%%%%%%%%%

In the continuum limit ($1/L\to 0)$, with $\epsilon_k=k=\alpha/L$, the characteristic equation can be expanded in powers of $1/L$. To leading order, ${\mathrm O}\left(L^{-2}\right)$, one obtains
\begin{eqnarray}
\cos\alpha&=&\frac{\left[\kappa_1-(-1)^Q\,\kappa_2\right]^2-\left[1+(-1)^Q\,\kappa_1\kappa_2\right]^2}{\left[\kappa_1-(-1)^Q\,\kappa_2\right]^2
+\left[1+(-1)^Q\,\kappa_1\kappa_2\right]^2}\,,\nonumber\\
\kappa_i&=&\frac{J_i}{h_{i1}\,h_{i2}}\qquad (i=1,2)\,,
\label{eA4}
\end{eqnarray}
where $\kappa_i$ is an effective bond interaction. This can be rewritten as
\be
\cot\left(\frac{\alpha}{2}\right)=\pm\frac{\kappa_1-(-1)^Q\,\kappa_2}{1+(-1)^Q\,\kappa_1\kappa_2}
=-\tan\left(\frac{\alpha}{2}-\frac{\pi}{2}\right)\,.
\label{eA5}
\ee
Introducing the angle $\theta_i=\arctan(1/\kappa_i)$ ($i=1,2$) leads to the relation 
\begin{eqnarray} 
\tan\left(\frac{\alpha}{2}-\frac{\pi}{2}\right)&=&\pm\frac{\tan\theta_1-(-1)^Q\,\tan\theta_2}{1+(-1)^Q\,\tan\theta_1\tan\theta_2}
\nonumber\\
&=&\pm\tan\left[\theta_1-(-1)^Q\,\theta_2\right]\,.
\label{eA6}
\end{eqnarray}
Thus the fermionic excitation energies take the form ~\cite{henkel87b,henkel89}
\be
\epsilon_r^\pm(Q)=\frac{2\pi}{L}\left(1/2\pm\Delta_Q+r\right)\,,\qquad Q=0,1\,,
\label{eA7}
\ee
where $r$ is an integer and
\be
\Delta_0=\left|\frac{\theta_1-\theta_2}{\pi}\right|\,,\quad\Delta_1=1-\frac{\theta_1+\theta_2}{\pi}\,.
\label{eA8}
\ee
The local magnetization exponent follows from the gap-exponent relation~\cite{cardy84a} 
\be
x_{12}=\frac{L}{2\pi}(E_\sigma-E_0)\,,
\label{eA9}
\ee
where $E_0=E_0(0)$ is the ground-state energy of ${\cal H}$ in equation~\eref{eA1} which belongs to the even sector of the fermionic Hamiltonian whereas $E_\sigma$, which is the first excited state of ${\cal H}$, belongs to the odd sector. Since these two states belong to different sectors the gap involves the difference $\Delta E$ between the ground-state energies in the two sectors and is given by
\be
E_\sigma-E_0=\Delta E+\epsilon_0(1)=E_0(1)-E_0(0)+\epsilon_0(1)\,,
\label{eA10}
\ee
with~\cite{henkel87b}
\be
\Delta E=\frac{2\pi}{L}\left[\frac{1}{2}\left(\Delta^2_1-\Delta^2_0\right)\right]\,,\quad
\epsilon_0(1)=1/2-\Delta_1\,.
\label{eA11}
\ee
Collecting these results, one finally obtains a simple expression for the local magnetization exponent
\be
x_{12}=\frac{2}{\pi^2}\arctan\left(\frac{1}{\kappa_1}\right)\arctan\left(\frac{1}{\kappa_2}\right)=\sqrt{x_1x_2}\,.
\label{eA12}
\ee

This result can be verified for the ``composite defect'' of figure~\ref{figA3}(b) where $\kappa_1=0$ (cut on the positive $u$-axis) and 
$\kappa_2=1$ (no perturbation on the negative $u$-axis). Then $x_1$ is the free surface exponent $x_{ms}=1/2$, $x_2$ the bulk exponent $x_m=1/8$ and equation~\eref{eA12} gives $x_{12}=1/4$. Let us now apply the conformal transformation $w=z^{\omega/\pi}$ to the critical upper half-plane of figure~\ref{figA3}(a). One obtains a corner with opening angle $\omega$ and corner exponents are related to surface exponents via $x_c=\pi x_{ms}/\omega$~\cite{cardy84b,barber84}. When $\omega=2\pi$, the transformed system is the full plane with a cut of figure~\ref{figA3}(b) so that $x_{12}=x_{ms}/2=1/4$.

The gap leading to the local energy density exponent is $E_\epsilon-E_0$, where $E_\epsilon$ is the lowest eigenstate of ${\cal H}$ with 
two fermions. Both states belong to the even sector so that, according to equation~\eref{eA7}, the local energy exponent, given by 
\be
\frac{L}{2\pi}(E_\epsilon-E_0)=\frac{L}{2\pi}\left[\epsilon_0^-(0)+\epsilon_0^+(0)\right]=1\,,
\label{eA13}
\ee
keeps its unperturbed value. This is necessary to keep a truly marginal behavior for the local magnetization when the defect strength is modified. Otherwise the marginality criterion would no longer be satisfied.

\end{document}